\documentclass[journal,twocolumn,10pt,twoside]{IEEEtran}
\usepackage[margin=1.6cm, centering]{geometry}
\usepackage{multirow}
\usepackage{booktabs}
\usepackage{lipsum}
\usepackage{amsfonts}
\usepackage{amsmath,cases,amssymb}
\usepackage{graphicx}
\usepackage{cite}
\usepackage{subfigure}
\usepackage{epsfig}
\usepackage{url}
\usepackage{algorithm}
\usepackage{algorithmic}
\usepackage{epstopdf}
\usepackage{balance}
\usepackage{color}

\newcommand{\ds}{\displaystyle}

\newcommand{\st}{{\mathrm{s.t.}}}

\newcommand{\clR}{{\cal R}}

\newcommand{\barR}{{\mathsf{\bar R}}}

\newcommand{\bfw}{\mathbf{w}}
\newcommand{\bfh}{\mathbf{h}}

\newcommand{\balpha}{\pmb{\alpha}}
\begin{document}
\title{Joint Fractional Time Allocation and Beamforming for Downlink Multiuser MISO Systems
\thanks{This work was supported in part by the National Research Foundation of Korea (NRF) grant funded by the Korean government (MSIT) (No. NRF-2017R1A5A1015596),  in part by Basic Science Research Program through the NRF funded by the Ministry of Education (No. 2017R1D1A1B03030436),     in part by the U.K. Royal Academy of Engineering Research Fellowship under Grant RF1415$\slash$14$\slash$22 and by the U.K. Engineering and Physical Sciences Research Council (EPSRC) under Grant EP/P019374/1, and in part by the KAUST Grant No. OSR-2016-CRG5-2958-0 and U.S. National Science Foundation Grant ECCS-1647198.  (\textit{Corresponding author: Oh-Soon Shin.}) }}

\author{
Van-Dinh Nguyen, Hoang Duong Tuan, Trung Q. Duong,  Oh-Soon Shin, and H. Vincent Poor 
\thanks{V.-D. Nguyen and O.-S. Shin are with the School of Electronic Engineering and the Department of ICMC Convergence Technology,
Soongsil University, Seoul 06978, Korea (e-mail: \{nguyenvandinh, osshin\}@ssu.ac.kr).}
\thanks{H. D. Tuan is with the Faculty of Engineering and Information Technology, University of Technology Sydney, Broadway, NSW 2007,
Australia (email: tuan.hoang@uts.edu.au).}
\thanks{T. Q. Duong is with the School of Electronics, Electrical Engineering and Computer Science, Queen's University Belfast, Belfast BT7 1NN, United Kingdom (e-mail: trung.q.duong@qub.ac.uk).}
\thanks{H. V. Poor is with the Department of Electrical Engineering, Princeton University, Princeton, NJ 08544 USA (e-mail: poor@princeton.edu).}
\vspace*{-0.5cm}}
\markboth{IEEE Communications Letters}%
{Nguyen \MakeLowercase{\textit{et al.}}: Joint Fractional Time Allocation and Beamforming for Downlink Multiuser MISO Systems}
\maketitle

\begin{abstract}It is well-known that the traditional transmit beamforming at a base station (BS) to manage interference in serving
multiple users is effective only when the number of users is less than the number of transmit antennas at the BS. Non-orthogonal multiple access (NOMA) can improve the throughput of users with poorer channel conditions by compromising their own privacy because other users with better channel conditions can decode the information of users in poorer channel state. NOMA still prefers that
the number of users is less than the number of antennas at the BS transmitter. This paper resolves such issues by allocating separate fractional time slots for serving the users with similar channel conditions. This enables the BS to serve more users within the time unit while the privacy of each user is preserved. The
fractional times and beamforming vectors are jointly optimized to maximize the system's throughput. An efficient path-following
algorithm, which invokes a simple convex quadratic program at each iteration, is proposed for the solution of this challenging optimization problem. Numerical results confirm its versatility.
\end{abstract}
\section{Introduction}\label{sec:intro}
In multi-user communication, the signal received by any user (UE) is a superposition of the desired information and the information intended for other UEs. Transmit beamforming at a base station (BS) plays a pivotal role in
focusing the energy of the desired signal and suppressing the multi-user interference to achieve better throughput at UEs \cite{ShangIT11,GSO}.
For effective beamforming, the number of UEs
usually does not exceed the number of transmit antennas. Using more transmit antennas and thus increasing the dimensionality of beamforming vectors can improve the UEs' throughput with reduced transmit power. However, allocating more transmit power under fixed number of transmit antennas still does not necessarily improve the UEs' throughput.

Non-orthogonal multiple access (NOMA) \cite{DSP16,DingMag17} has been introduced to improve
the UEs' throughput  by allowing UEs with better channel conditions to access
and decode the signals, which are intended for the UEs with poorer channel conditions. In other words, the UEs with
poorer channel conditions can achieve higher throughput by compromising their information privacy in NOMA \cite{DingTWC16}.
This privacy compromise for the UEs with poor channel conditions is unavoidable in NOMA. Moreover, by restricting all beamforming vectors in the same space, beamforming in NOMA still needs that the dimension of this space, which is
equal to the number of transmit antennas, should not be less than the number of UEs to allow the suppression of multiuser interference \cite{Dietal17}.

It is noteworthy that the UEs with good channel conditions may need only a fraction of the time unit to achieve their needed throughput.
Therefore, by serving them only per fractional time, the BS still has the remaining time room to serve
the UEs with poor channel conditions. In this way, the information privacy for each UE is preserved because all UEs are allowed to
decode their own information only. More importantly, the number of UEs supported at the same fractional
time is effectively reduced. Thus, the BS will not need more transmit antennas to suppress the multi-user interference.
In this letter, the fractional time allocation to UEs with similar channel conditions and beamforming are enhanced
for improving the system's throughput and meeting the UEs' quality-of-service (QoS) in terms of the throughput requirement.
This problem is mathematically modelled by a highly nonconvex optimization problem, for which a path-following computational procedure
of low complexity is then developed for its computation. Finally, the numerical examples are provided to demonstrate the
advantage of the proposed optimization scheme.

\textit{Notation.} We use bold upper-case letters for matrices, bold lower-case letters for column vectors, lower-case letters for scalars. $\Re\{x\}$ denotes the real part of a complex number $x$. The notation $(\cdot)^{H}$ stands for the Hermitian transpose. $\mathbf{x}\sim\mathcal{CN}(\boldsymbol{\eta},\boldsymbol{Z})$ means that $\mathbf{x}$ is a random vector following a complex circular Gaussian distribution with mean  $\boldsymbol{\eta}$ and covariance matrix $\boldsymbol{Z}$.
\section{System Model and Problem Formulation}\label{sec:sys_model}
\begin{figure}[t]
\centering
\includegraphics[width=0.28\textwidth,trim={0cm 0.0cm -0cm -0cm}]{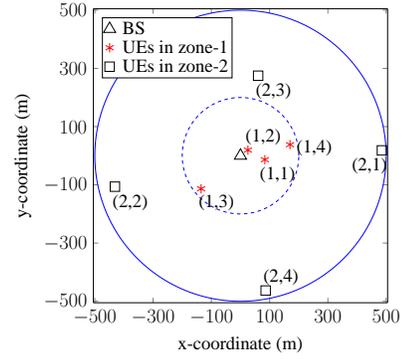}
\caption{Scenario with $K=4$.}
\label{fig:Layout}
\end{figure}

Consider a multiuser downlink system over a given frequency band with
a BS equipped with $N_t > 1$ antennas in serving $2K$  single-antenna UEs as illustrated by Fig. \ref{fig:Layout}.
 There are $K$ UEs $(1,k)$, $k=1,\dots K$, which are located in a zone nearer to the BS, called by zone-1,
and $K$  UEs $(2,k)$, $k=1,\dots, K$, which are located in a zone farer from the BS, called by zone-2.
Denote by $\mathcal{K} \triangleq\{1,2,\dots, K\}$ and  $\mathcal{M}=\{1,2\}\times \mathcal{K}$.
Within one time unit, BS uses the fraction time (FT) $\tau_1:=\tau\;(0 < \tau < 1)$ to serve UEs $(1,k)$
and uses the remaining FT $\tau_2:=(1 - \tau)$ to serve UEs $(2,k)$.

The BS deploys a transmit beamformer $\bfw_{i,k}\in\mathbb{C}^{N_t\times 1}$ to deliver the information signal $x_{i,k}$ with $\mathbb{E}\{|x_{i,k}|^2 \} =1$ to UE $(i,k)$.
Let $\bfh_{i,k}\in\mathbb{C}^{N_t\times 1}$ be the channel vector from the BS to UE $(i,k)$, which is assumed to follow frequency flat fading with the effects of both large-scale pathloss and small-scale fading counted. The complex baseband signal received by UE $(i,k)$ can be expressed as
\[
y_{i,k} = \bfh_{i,k}^H\bfw_{i,k}x_{i,k} + \sum_{j\in\mathcal{K}\setminus \{k\}}\bfh_{i,k}^H\bfw_{i,j}x_{i,j}  + n_{i,k}
\]
where the first term is the desired signal, the second term is the multi-user interference, and the third term
$n_{i,k}\sim\mathcal{CN}(0,\sigma^2_{i,k})$ is additive  noise.
{\color{black}For $\bfw_i\triangleq (\bfw_{i,k})_{k\in{\cal K}}$,
the throughput at UE $(i,k)$ is
\[
\clR_{i,k}\bigl(\bfw_i,\tau_i \bigr) = \tau_i\ln\Bigl(1+\frac{ |\bfh_{i,k}^H\bfw_{i,k}|^2}{\sum_{j\in\mathcal{K}\setminus \{k\}}|\bfh_{i,k}^H\bfw_{i,j}|^2 + \sigma^2_{i,k}}\Bigl)
\]
which by  \cite{WLP} can be equivalently reformulated by
\begin{equation}\label{frate}
\clR_{i,k}\bigl(\bfw_i,\tau_i \bigr) = \tau_i\ln\Bigl(1+\frac{ (\Re\{\bfh_{i,k}^H\bfw_{i,k}\})^2}{\sum_{j\in\mathcal{K}\setminus \{k\}}|\bfh_{i,k}^H\bfw_{i,j}|^2 + \sigma^2_{i,k}}\Bigl)
\end{equation}
under the additional condition
\begin{equation}\label{ad}
\Re\{\bfh_{i,k}^H\bfw_{i,k}\} \geq 0, (i,k)\in {\cal M}.
\end{equation}
}
The main advantage of this FT-based beamforming scheme is that there is no inter-zone interference in (\ref{frate})
that is in contrast with the
conventional scheme to concurrently serve all UEs, under which the throughput at UE $(i,k)$ is
\begin{equation}\label{crate}
\clR_{i,k}^{'}(\bfw) = \ln\Bigl(1+ \frac{(\Re\{\bfh_{i,k}^H\bfw_{i,k}\})^2}{\sum_{(i,j)\in\mathcal{M}\setminus \{(i,k)\}}|\bfh_{i,k}^H\bfw_{i,j}|^2 + \sigma^2_{i,k}}\Bigl)
\end{equation}
with the full inter-zone interference. Here and in the sequence $\bfw\triangleq (\bfw_1, \bfw_2)$ and $\boldsymbol{\tau}\triangleq(\tau_1,\tau_2)$.

We are interested in the following problem of jointly designing FT $(\tau_1,\tau_2)$ and the beamformers $(\bfw_1, \bfw_2)$ to maximize the system sum throughput (ST):
\begin{subequations}\label{eq:SRoptimi}
    \begin{eqnarray}
        &\underset{\bfw,\boldsymbol{\tau}}{\max}& \sum_{(i,k)\in{\cal M}} \clR_{i,k}(\bfw_i,\tau_i) \quad \st \quad (\ref{ad}), \label{eq:SRoptimi:a}\\
        &&\clR_{i,k}\bigl(\bfw_i,\tau_i  \bigr) \geq \barR_{i,k},\  \forall(i,k)\in\mathcal{M}, \label{eq:SRoptimi:b}\\
        &&\tau_1\|\bfw_{1}\|^2 + \tau_2\|\bfw_{2}\|^2 \leq P_{bs}^{\max},\label{eq:SRoptimi:d}\\
				&& \tau_1\geq 0, \tau_2\geq 0, \tau_1+\tau_2\leq 1. \label{eq:SRoptimi:e}
         \end{eqnarray}
    \end{subequations}
Here $\barR_{i,k}$ sets a minimum throughput requirement for UE $(i,k)$ and $P_{bs}^{\max}$ is a given power budget.
Since $\clR_{i,k}(\bfw_i,\tau_i)$ is a nonconcave function, the optimization problem  (\ref{eq:SRoptimi}) is regarded
as a highly nonconvex optimization problem, for which finding a feasible point is already computationally difficult. The next
section is devoted to a computational path-following procedure for its solution.

\section{Convex Quadratic-based Path-following method}\label{sec:CQBI}
The most important step is to provide an effective lower bounding approximation for the function
$\clR_{i,k}(\bfw_i,\tau_i)$ defined by (\ref{frate}) to facilitate
a path-following computational procedure  of the  problem  (\ref{eq:SRoptimi}).
We use the  variable changes $\balpha\triangleq (\alpha_1, \alpha_2)=(1/\tau_1,1/\tau_2)$,
which satisfy the following convex constraint:
\begin{equation}\label{eq:changevariables:b}
  1/\alpha_1+1/\alpha_2 \leq 1, \alpha_1>0, \alpha_2>0.
\end{equation}
The  problem \eqref{eq:SRoptimi} can be equivalently expressed as
\begin{subequations}\label{eq:SRMequi}
    \begin{eqnarray}
        &\underset{\bfw, \balpha}{\max}&  \Phi(\bfw,\balpha)\triangleq \sum_{(i,k)\in{\cal M}}
        \clR_{i,k}(\bfw_i,1/\alpha_i)\quad \st\quad (\ref{ad}), \eqref{eq:changevariables:b},\label{eq:SRMequi:a}\qquad\\
        &&\clR_{i,k}(\bfw_i,1/\alpha_i)  \geq \barR_{i,k},\ \forall(i,k)\in{\cal M},\label{eq:SRMequi:b}\\
         &&(1-1/\alpha_2)\|\bfw_{1}\|^2 + \|\bfw_{2}\|^2/\alpha_2 \leq P_{bs}^{\max}.\label{eq:SRMequi:d}
         \end{eqnarray}
    \end{subequations}
At a feasible point $(\bfw^{(\kappa)},\balpha^{(\kappa)})$, define
$x_{i,k}^{(\kappa)}\triangleq \Re\{\bfh_{i,k}^H\bfw_{i,k}^{(\kappa)}\}>0$,
$y_{i,k}^{(\kappa)}\triangleq\sum_{j\in\mathcal{K}\setminus \{k\}}|\bfh_{i,k}^H\bfw_{i,j}^{(\kappa)} |^2 + \sigma^2_{i,k} $,
$d_{i,k}^{(\kappa)}  \triangleq (x_{i,k}^{(\kappa)})^2/y_{i,k}^{(\kappa)}>0$,
$a^{(\kappa)}_{i,k}\triangleq 2\clR_{i,k}(\bfw_i^{(\kappa)},1/\alpha_i^{(\kappa)})+d_{i,k}^{(\kappa)}/\alpha_i^{(\kappa)}(d_{i,k}^{(\kappa)}+1)>0$,
$b^{(\kappa)}_{i,k}\triangleq (d_{i,k}^{(\kappa)})^2/\alpha_i^{(\kappa)}(d_{i,k}^{(\kappa)}+1)>0$, and
$c_{i,k}^{(\kappa)}\triangleq \clR_{i,k}(\bfw_i^{(\kappa)},1/(\alpha_i^{(\kappa)})^2)>0$.
It follows from the inequality \eqref{inq1} in the appendix
that
\[
\clR_{i,k}(\bfw_i,1/\alpha_i)\geq \clR_{i,k}^{(\kappa)}(\bfw_i,\alpha_i)
\]
over the trust region
\begin{IEEEeqnarray}{rCl}
2\Re\{\bfh_{i,k}^H\bfw_{i,k}\} - \Re\{\bfh_{i,k}^H\bfw_{i,k}^{(\kappa)}\} > 0,\  \forall(i,k) \in\mathcal{M},
\label{eq:Rate1ktrust}
 \end{IEEEeqnarray}
for the concave function
\[
\clR_{i,k}^{(\kappa)}(\bfw_i,\alpha_i)\triangleq
a^{(\kappa)}_{i,k}-
b^{(\kappa)}_{i,k}\frac{\sum_{j\in\mathcal{K}\setminus \{k\}}|\bfh_{i,k}^H\bfw_{i,j}|^2 + \sigma^2_{i,k} }{x_{i,k}^{(\kappa)}(2\Re\{\bfh_{i,k}^H\bfw_{i,k}\}-x_{i,k}^{(\kappa)})}
- c_{i,k}^{(\kappa)}\alpha_i.
\]
Next, due to the convexity of function $\|\bfw_{1}\|^2/\alpha_2$, it is true that
$ \|\bfw_{1}\|^2/\alpha_2\geq 2\Re\{(\bfw_{1}^{(\kappa)})^H\bfw_{1}\}/\alpha_2^{(\kappa)}- (\|\bfw_{1}^{(\kappa)}\|^2/(\alpha_2^{(\kappa)})^2)\alpha_2$.
An inner convex approximation of nonconvex constraint \eqref{eq:SRMequi:d} is then given by
\begin{eqnarray}
 \|\bfw_{1}\|^2 + \|\bfw_{2}\|^2/\alpha_2 - 2\Re\{(\bfw_{1}^{(\kappa)})^H\bfw_{1}\}/\alpha_2^{(\kappa)}&\nonumber\\
 + (\|\bfw_{1}^{(\kappa)}\|^2/(\alpha_2^{(\kappa)})^2)\alpha_2 &\leq P_{bs}^{\max}.\label{eq:SRMequi:d2}
 \end{eqnarray}
Initialized by a feasible point $(\bfw^{(0)},\balpha^{(0)})$ for (\ref{eq:SRMequi}), the following convex quadratic program (QP)
is solved at the $\kappa$-th iteration to generate the next feasible point $(\bfw^{(\kappa+1)},\balpha^{(\kappa+1)})$:
\begin{eqnarray}
\ds\max_{\bfw,\balpha}\  \Phi^{(\kappa)}(\bfw,\balpha)\triangleq \sum_{(i,k)\in{\cal M}}\clR_{i,k}^{(\kappa)}(\bfw_i,\alpha_i)\quad\st\nonumber\\
 \clR_{i,k}^{(\kappa)}(\bfw_i,\alpha_i)\geq \barR_{i,k},\ \forall(i,k) \in\mathcal{M}, (\ref{ad}),
\eqref{eq:changevariables:b}, \eqref{eq:Rate1ktrust},  \eqref{eq:SRMequi:d2}.  \label{QPproblem}
\end{eqnarray}
As problem \eqref{QPproblem} involves
$m=2(3K+1)$ quadratic and linear constraints, and $n=2(KN_t  +  1)$ real decision variables, its
computational complexity is $\mathcal{O}(n^2m^{2.5}+m^{3.5})$.

Note that $\Phi(\bfw,\balpha)\geq \Phi^{(\kappa)}(\bfw,\balpha)$ $\forall (\bfw,\balpha)$, and
$\Phi(\bfw^{(\kappa)},\balpha^{(\kappa)})=\Phi^{(\kappa)}(\bfw^{(\kappa)},\balpha^{(\kappa)})$. Moreover,
$\Phi^{(\kappa)}(\bfw^{(\kappa+1)},\balpha^{(\kappa+1)})>\Phi^{(\kappa)}(\bfw^{(\kappa)},\balpha^{(\kappa)})$
whenever $(\bfw^{(\kappa+1)},\balpha^{(\kappa+1)})\neq (\bfw^{(\kappa)},\balpha^{(\kappa)})$ because the former and
the latter, respectively, are the optimal solution and  feasible point for (\ref{QPproblem}). Therefore,
$\Phi(\bfw^{(\kappa+1)},\balpha^{(\kappa+1)})\geq \Phi^{(\kappa)}(\bfw^{(\kappa+1)},\balpha^{(\kappa+1)})>\Phi^{(\kappa)}(\bfw^{(\kappa)},\balpha^{(\kappa)})=
\Phi(\bfw^{(\kappa)},\balpha^{(\kappa)})$, showing that $(\bfw^{(\kappa+1)},\balpha^{(\kappa+1)})$ is a better feasible point
than $(\bfw^{(\kappa)},\balpha^{(\kappa)})$ for (\ref{eq:SRMequi}). The sequence $\{(\bfw^{(\kappa)},\balpha^{(\kappa)})\}$
of improved feasible points for (\ref{eq:SRMequi}) thus converges at least to a locally optimal solution satisfying the
Karush-Kuh-Tucker conditions \cite{MW78}. We summarize the proposed QP-based
path-following procedure  in Algorithm~\ref{alg_SCALE_FW}.
\begin{algorithm}[t]
\begin{algorithmic}[1]
\protect\caption{QP-based path-following algorithm for ST maximization problem \eqref{eq:SRoptimi}}
\label{alg_SCALE_FW}
\global\long\def\algorithmicrequire{\textbf{Initialization:}}
\REQUIRE  Iterate  \eqref{qosconvex} for an initial feasible point $(\bfw^{(0)},\balpha^{(0)})$. Set $\kappa:=0$
\REPEAT
\STATE Solve convex quadratic program \eqref{QPproblem} to obtain the optimal solution: $(\bfw^{(\kappa+1)},\balpha^{(\kappa+1)})$.
\STATE Set $\kappa:=\kappa+1.$
\UNTIL Convergence
\end{algorithmic} \end{algorithm}

\textit{Generation of an initial point.}: Initialized from a feasible point $(\bfw^{(0)},\balpha^{(0)})$ for constraints
 \eqref{eq:changevariables:b} and (\ref{eq:SRMequi:d2}), we  iterate the convex program
\begin{equation}\label{qosconvex}
\max_{\bfw, \balpha} \min_{(i,k) \in\mathcal{M}} \clR_{i,k}^{(\kappa)}(\bfw_i,\alpha_i)/\bar{\mathsf{R}}_{i,k}\quad
\st\quad (\ref{ad}), \eqref{eq:changevariables:b}, \eqref{eq:Rate1ktrust},\eqref{eq:SRMequi:d2}
\end{equation}
till reaching
$\min_{(i,k)\in{\cal M}} \clR_{i,k}^{(\kappa)}(\bfw_i^{(\kappa+1)},\alpha_i^{(\kappa+1)})/\bar{\mathsf{R}}_{i,k} \geq 1$
to make $(\bfw^{(\kappa+1)},\balpha^{(\kappa+1)})$ feasible for (\ref{eq:SRMequi}) and thus usable as an initial feasible
point for implementing Algorithm \ref{alg_SCALE_FW}.
\section{Numerical Results}\label{sec:simulation}
Monte Carlo simulations have been implemented to evaluate the performance of the proposed algorithm for $K=4$ ($8$ UEs)
and $N_t=5$ per the scenario  in Fig.~\ref{fig:Layout}. The channel vector $\mathbf{h}_{i,k}$ between the BS and
UE $(i,k)$ at a distance $d_{i,k}$ (in kilometres) is generated  as $\mathbf{h}_{i,k}=\sqrt{10^{-\sigma_{\mathsf{PL}}/10}}\tilde{\mathbf{h}}_{i,k}$, where $\sigma_{\mathsf{PL}}$ is the path loss (PL) in dB and  $\tilde{\mathbf{h}}_{i,k}\sim\mathcal{CN}(0,\mathbf{I}_{N_t})$ represents small-scale effects.
The other parameters are given by Table \ref{parameter}.   Without loss of generality, $\bar{\mathsf{R}}_{i,k}\equiv \bar{\mathsf{R}}$ is set. The numerical results are obtained using the parser YALMIP \cite{L04}.

We compare the  performance of the proposed FT-based beamforming scheme with  five other beamforming schemes:
$(i)$  ``Conventional DL,'' under which the problem of ST maximization is formulated similarly as:
 $\max_{\bfw}\sum_{(i,k)\in {\cal M}} \clR_{i,k}^{'}(\bfw)\ \st\ \clR_{i,k}^{'}(\bfw)\geq \barR_{i,k}, (i,k)\in{\cal M},
 \|\bfw_1\|^2+\|\bfw_2\|^2\leq P_{bs}^{\max}$ under the definition (\ref{crate});
$(ii)$ ``NOMA'':  {\color{black}each UE in zone-1 is  paired with an UE in zone-2 according to the clustering algorithm
in \cite{Kietal13}  to create a virtual cluster.} In each cluster, both UEs decode the signal intended for the UE in zone-2
   and then the UE in zone-1 processes successive interference cancellation (SIC) to cancel the interference of the UE in zone-2 in decoding its own signal;
$(iii)$ ``FT + NOMA in both zones'': under FT, NOMA is zone-wide adopted;
$(iv)$ ``FT + NOMA in zone-1'': under FT, NOMA is adopted only in zone-1; and
$(v)$ ``FT + NOMA in zone-2'': under FT, NOMA is adopted only in zone-2. {\color{black}The reader is referred to \cite[Sec. V]{Dietal17}
for beamforming under NOMA, which is used in these five schemes. {\color{black}The
computational complexity of each iteration in NOMA is similar to that of (\ref{QPproblem}).}
Note that
the performance of NOMA-based beamforming can be improved by involving more UEs in virtual clusters \cite[Sec VI]{Dietal17} but the UEs' privacy is more compromised.}
\begin{table}[t]
\caption{Simulation Parameters}
	\label{parameter}
	\centering
		\begin{tabular}{l|l}
		\hline
				Parameters & Value \\
		\hline\hline
		  	Noise power density & -174 [dBm/Hz] \\
				Path loss from the BS to UE $(i,k)$, $\sigma_{\mathsf{PL}}$   & 128.1 + 37.6$\log_{10}(d_{i,k})$ [dB]\\
				Radius of  cell &  500 [m]\\
				Coverage of zone-1 UEs  & 200 [m]\\
				Distance between the BS and nearest user & $>$ 10 [m]\\
					\hline		   				
		\end{tabular}
\end{table}	
On average, Algorithm~\ref{alg_SCALE_FW} requires about $10$ iterations for convergence.

\begin{figure}
    \begin{center}
    \begin{subfigure}[$\bar{\mathsf{R}} = 0 $ bps/Hz.]{
        \includegraphics[width=0.32\textwidth]{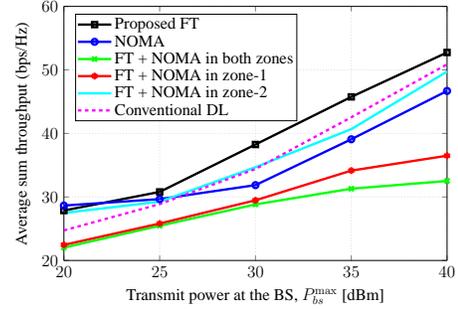}}
    		\label{fig:SRMa}
				\end{subfigure}
				 \begin{subfigure}[$\bar{\mathsf{R}} = 1 $ bps/Hz.]{
        \includegraphics[width=0.32\textwidth]{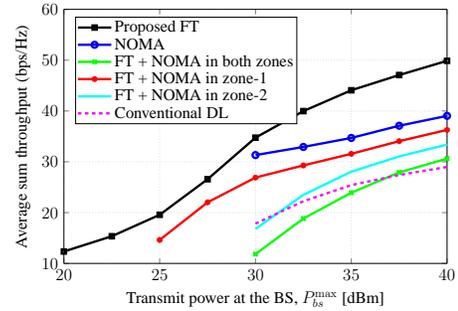}}
        \label{fig:SRMb}
    \end{subfigure}
	  \caption{Average sum throughput versus $P_{bs}^{\max}$.}\label{fig:SRvsPmax}
\end{center}
\end{figure}

Fig.~\ref{fig:SRvsPmax} plots the average achievable ST versus the transmit power $P_{bs}^{\max}$ for $N_t = 5$.   For $\bar{\mathsf{R}} = 0 $ bps/Hz shown in Fig.~\ref{fig:SRvsPmax}(a), one can see that  the  ST of
the proposed FT-based beamforming is higher than that achieved by the other schemes
in the high transmit power region. On the other
hand, the conventional DL outperforms NOMA and FT+NOMA schemes for high transmit power
by making the ST concentrated at those UEs with good channel conditions.
Apparently, NOMA does not look efficient when there is no UEs' QoS requirement.
In Fig.~\ref{fig:SRvsPmax}(b) with  $\bar{\mathsf{R}} = 1 $ bps/Hz, the conventional DL performs worse than
NOMA at high $P_{bs}^{\max}$. The low STs of FT+NOMA-based schemes are probably attributed to the fact that  NOMA is more efficient by exploiting their channel differentiation. Increasing $P_{bs}^{\max}$ leads to a remarkable gain in ST
by the proposed  FT compared with the other schemes. In addition, the proposed FT is feasible in all range of $P_{bs}^{\max}$ while the other schemes cannot offer such high QoS to UEs at low $P_{bs}^{\max}$.

\begin{figure}
\centering
\includegraphics[width=0.32\textwidth]{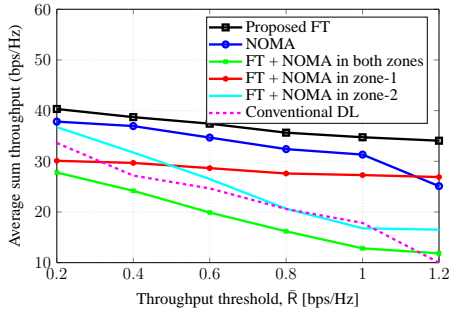}
	  \caption{Average sum throughput  versus $\bar{\mathsf{R}}$ for $P_{bs}^{\max} = 30$ dBm.}\label{fig:SRvsRth}
\end{figure}

The plot of the ST  versus QoS requirement threshold $\bar{\mathsf{R}}\in[0.2,\;1.2]$ bps/Hz is illustrated in Fig.~\ref{fig:SRvsRth}. We can observe that the proposed FT-based beamforming  performs quite well and only slightly degrades when $\bar{\mathsf{R}}$ increases. The performance gap between the proposed FT-based beamforming and other schemes  substantially increases by increasing $\bar{\mathsf{R}}$. It is expected because the proposed
FT-based beamforming can tune the power allocation in meeting zone-2 UEs' QoS requirements without causing interference to  the zone-1's UEs.

Next, we look for the max-min UE throughput optimization problem
\begin{equation}\label{eq:maxminoptimi}
\max_{\bfw, \boldsymbol{\tau}}\min_{(i,k) \in\mathcal{M}}\clR_{i,k}(\bfw_i,\tau_i)\quad
\st\quad \eqref{eq:SRoptimi:d}, \eqref{eq:SRoptimi:e}
    \end{equation}
which can be addressed similarly by solving the convex program (\ref{qosconvex})
(with $\bar{\mathsf{R}}_{i,k}\equiv 1$) at each iteration.
\begin{figure}
    \centering
        \includegraphics[width=0.32\textwidth]{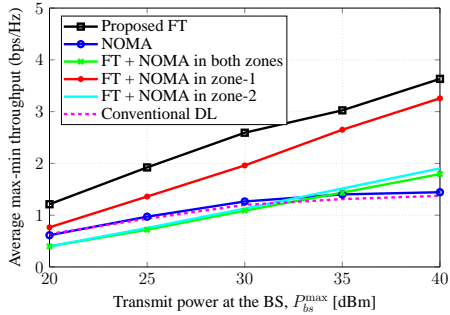}
     \caption{Average max-min throughput versus $P_{bs}^{\max}$.}
		\label{fig:MaxminPbs}
\end{figure}

Fig.~\ref{fig:MaxminPbs} plots the average UEs' worst throughput as a function of $P_{bs}^{\max}$.
It shows that the worst throughput achieved by
  the conventional DL and FT + NOMA-based schemes is saturated once $P_{bs}^{\max}$ is beyond a threshold. It also
  reveals that the proposed FT-based beamforming
   and FT + NOMA in zone-1 achieve worst throughput that is higher than that achieved by others schemes.

\section{Conclusions}\label{sec:conclusion}
The paper has proposed a fractional time-based beamforming scheme at a base station to serve two groups of users, which
is able to improve the network throughput while preserving the information privacy for the users. Accordingly,
a path-following computational procedure for the joint design of fractional times and
beamforming vectors  to maximize the network throughput has been developed. Extensive simulations  have been
provided to demonstrate the superior performance of the proposed scheme over the exiting schemes.
\section*{Appendix} \label{Appendix:A}
For all $ x > 0$, $\bar{x} > 0$, $y>0$, $\bar{y}>0$, $t>0$, and $\bar{t}>0$, it is true that
\begin{eqnarray}
\ds\frac{\ln(1+x^2/y)}{t}&\geq&\ds a - b\frac{y}{x^2} - ct\label{ing1a}\\
&\geq&\ds a - b\frac{y}{\bar{x}(2x-\bar{x})} - ct\label{inq1}
\end{eqnarray}
over the trust region
\begin{equation}\label{trust2}
2x-\bar{x}>0,
\end{equation} where
$a = 2 [\ln(1+d)]/\bar{t} +
d/\bar{t}(d+1) > 0, b=d^2/\bar{t}(d+1) > 0,\\
c = [\ln(1+d)]/\bar{t}^2> 0, d=\bar{x}^2/\bar{y}.$ 
Inequality (\ref{ing1a}) follows
from \cite{Ngetal17} while inequality (\ref{inq1}) is obtained by using $x^2\geq \bar{x}(2x-\bar{x})>0$ over the
trust region (\ref{trust2}).
\bibliographystyle{IEEEtran}
\balance
\bibliography{Journal}
\end{document}